\documentclass[unnumsec,webpdf,modern,large]{mam-authoring-template}%

\usepackage[utf8]{inputenc}

\usepackage{amsfonts,amsmath,amssymb,bm}
\usepackage{systeme}
\usepackage{econometrics}
\usepackage{graphics}
\usepackage{dsfont}
\usepackage{graphicx}
\usepackage{siunitx}
\usepackage[default]{lato}
\usepackage[T1]{fontenc}
\usepackage{cancel}
\usepackage{float}
\usepackage{booktabs}
\usepackage{pdfpages}
\usepackage{hyperref}
\usepackage{graphicx}
\usepackage{amsmath}
\usepackage{graphicx}
\usepackage{enumerate}
\usepackage{amsfonts}
\usepackage{url} 
\usepackage{kpfonts}

\newcommand{\bX}{\boldsymbol{X}}
\newcommand{\bH}{\boldsymbol{H}}
\newcommand{\bS}{\boldsymbol{S}}
\newcommand{\bT}{\boldsymbol{T}}

\newcommand{\bD}{\boldsymbol{D}}
\newcommand{\bx}{\boldsymbol{x}}

\newcommand{\btheta}{\boldsymbol{\theta}}

\newcommand{\MSES}{\mathrm{MSE}}

\newcommand{\FI}{\mathrm{FI}}
\newcommand{\ODAL}{\mathrm{ODAL}}
\newcommand{\pool}{\mathrm{pool}}

\begin{document}

\journaltitle{arXiv}
\copyrightyear{2026}


\title[An Accurate and Single-Communication Federated Inference Algorithm]{An Accurate and Privacy-Preserving Single-Communication Federated Inference Algorithm}

\author[1]{Laura Montagnani}
\author[1,2]{Anthony CC Coolen}
\author[3,$\ast$]{Marianne A Jonker}

\authormark{Montagnani et al.}

\address[1]{\orgdiv{Donders Institute, Faculty of Science}, \orgname{Radboud University}, \orgaddress{\state{Nijmegen}, \country{Netherlands}}}

\address[2]{\orgdiv{Saddle Point Science Europe, Mercator Science Park}, \orgaddress{\state{Nijmegen}, \country{Netherlands}}}

\address[3]{\orgdiv{Research Institute for Medical Innovation, Science department IQ Health, Section Biostatistics}, \orgname{Radboud university medical center}, \orgaddress{\state{Nijmegen}, \country{Netherlands}}}


\corresp[$\ast$]{Corresponding author: Marianne Jonker,~~ \href{email:email-id.com}{marianne.jonker@radboudumc.nl}}

\received{Date}{0}{Year}
\revised{Date}{0}{Year}
\accepted{Date}{0}{Year}

\abstract{{\bf Objective:}
Joint analyses across multiple institutions are increasingly important in biomedical and epidemiological research, particularly for rare diseases where datasets are typical small. However, privacy regulations and institutional policies often prevent the sharing of individual-level patient data. In this paper we present an accurate and single-communication federated inference algorithm.\\
{\bf Methods:}
Single-communication federated inference enables statistical analyses through a single exchange of summary statistics between participating centers and a coordinating server, preserving privacy while reducing communication and computational costs compared with iterative federated learning. In this paper, we extend a recently proposed single-communication federated inference strategy that is based on second-order Taylor expansions by using third-order expansions to better approximate local log-likelihood functions. The proposed method is evaluated through simulation studies based on real data and compared with existing federated inference strategies.\\
{\bf Results:}
The simulation studies assess the performance of the proposed method, with a particular focus on scenarios involving small local sample sizes, where quadratic approximations may fail to capture skewness and other higher-order characteristics of the log-likelihood function.\\
{\bf Discussion and Conclusion:}
Simulation studies demonstrate that incorporating higher-order information of the log-likelihood function  improves the accuracy while preserving the privacy, communication efficiency, and scalability required for collaborative biomedical and epidemiological research.}

\maketitle

\section{Introduction}
In many real-world research domains, particularly in biomedical and epidemiological research, joint data analysis between multiple institutions is becoming increasingly important. Pooled datasets from different centers allow researchers to study rare diseases \cite{Lassche}, for which only small datasets are available at individual institutions \cite{Mitani, Monaco}, improve statistical accuracy, and increase the generalizability of findings across patient populations. However, strict privacy regulations and institutional governance policies often complicate the sharing of individual-level patient data between institutions. As a result, many potentially valuable multi-center analyses cannot be performed using traditional statistical approaches in which all data are pooled in a single location. Therefore, the scientific challenge is to extract and combine the analyses results from different non-overlapping datasets to obtain the estimates that would have been found if the datasets had been pooled together.

To address this challenge, Federated Learning (FL) and analysis methods have been developed that allow privacy-preserving inference without combining datasets \cite{Kommusaar}. Many of these proposed methods are cyclic by design, requiring multiple iteration rounds between centers and the coordinating server \cite{Mcmahan, Rieke}, which increases computational and organizational costs. Moreover, the overall efficiency of cyclic federated approaches is constrained by the slowest participating center and when a new center joins the collaboration, additional communication rounds involving all participating centers become necessary, further increasing the complexity and costs of the process. 

In order to handle these limitations, so-called single-communication or one-shot federated statistical inference strategies have been proposed \cite{Duan, Jordan, Jonker2024, Luo, Xiong}. These methods aim to perform statistical analysis based on summary statistics computed from the data in the participating centers, requiring only a single communication between each center and the coordinating server while preserving patient privacy. Such one-shot federated methods demand far less coordination across centers, are easier to implement, are cheaper and are better suited for multi-institutional medical research settings. 

A promising recent development in this area is the Bayesian Federated Inference (BFI) framework \cite{Jonker2024, Jonker2025, Massa, Pazira}. The core idea is to perform Bayesian analyses locally at each site, producing approximations of the posterior parameter distributions, and then aggregate these approximated local posteriors into a global inference. The existing BFI methodology (BFI2) relies on second-order Taylor expansions (i.e., quadratic approximations) of the local log-posterior densities \cite{Jonker2024}. In practice, each center computes a quadratic approximation of the local log-posterior density near the local maximum a posteriori (MAP) estimate, which is shared with the coordinating server and combined with the quadratic approximations of the other centers to obtain an approximation of the pooled-data posterior density. The value that maximizes this approximate pooled-data density is defined as the the BFI2 estimate. This procedure is already being used in multiple ongoing collaborative research projects, which demonstrates its practical feasibility and highlights its potential to enable privacy-preserving collaboration between institutions. 

At the same time, these applications reveal an important methodological limitation: many real-world biomedical datasets are characterized by small sample sizes, what is typical for rare diseases. In such settings, the quadratic approximation may not hold sufficiently well since local log-posterior densities may exhibit skewness or other characteristics that cannot be captured by a second order approximation. As a consequence, the combined federated posterior distribution may deviate from the posterior that would have been obtained if the data had been combined. Improving the accuracy of federated inference in such settings is an important methodological challenge, which will be discussed in this paper. 

Indeed, the goal is to move beyond quadratic approximations by incorporating higher-order terms. By including third-order Taylor terms (which gives a cubic approximation), it may be possible to capture skewness and other features that are ignored by the previous approach. The main overarching goal of this paper is to explore whether such higher-order approximations can produce more accurate single-communication federated inference for non-overlapping datasets.  

This paper is organized as follows. First multiple federated inference strategies, including the algorithm that is based on a higher-order approximation, are described in the Methods section. Next, the performance of the multiple algorithms are compared by means of simulation studies based on real data. The paper ends with a discussion.

\section{Methods}
\subsection{Notation and setting}
\begin{figure}
\centering
\includegraphics[scale=0.28]{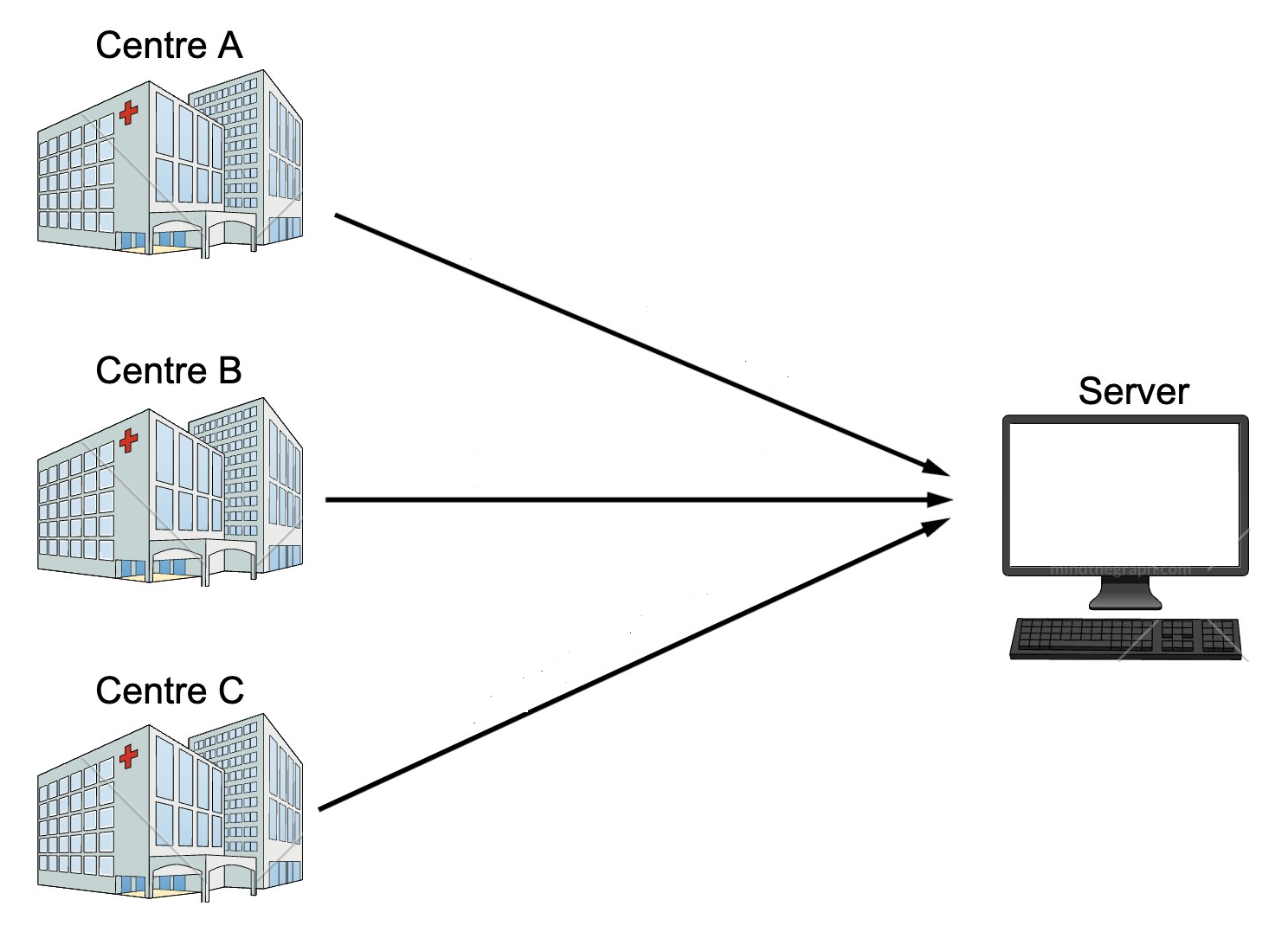}
\caption{Multiple medical centers perform local analyses and send summary statistics to a coordinating server. The server aggregates the summaries to produce inference results that approximate those obtained from pooled individual-level data. }
\label{fig: BFI}
\end{figure}

Suppose that data from $K$ different medical centers are available, denoted as $\bD_1, \ldots, \bD_K$ and with sample sizes $n_1,\ldots, n_K$. The hypothetical pooled dataset is denoted as $\bD = \cup_{k=1}^K \bD_k$ with sample size $n=\sum_{k=1}^K n_k$. We assume that all individual data within centers and across centers are independent and come from a parametric distribution which is known up to a finite dimensional parameter $\btheta \in {\mathbb R}^d$. Our aim is to estimate $\btheta$ using all available data without sharing individual-level observations across centers, while obtaining an estimator that closely approximates the one that would have been obtained had the data been pooled for a centralized analysis. To achieve this, each medical center performs the analysis on its local dataset and transmits only selected summary statistics to a coordinating server, where these summaries are combined to produce the final estimate (see Figure~\ref{fig: BFI}).

First the one-shot Bayesian Federated Inference methodology that was proposed in \cite{Jonker2024} is described. Next, an improved estimator is derived. Because one of the aims is to compare the proposed methods to the algorithm ODAL2 \cite{Duan}, which is defined in a frequentist framework, we decided to describe all methods in a frequentist way. Translation to a Bayesian version is straightforward.  

Let $\ell(\btheta)$ be the log-likelihood function  based on the hypothetical pooled dataset $\bD$, and let $\ell_k(\btheta)$ denote the log-likelihood function corresponding to the local dataset $\bD_k$ at center $k$. Under the assumption that observations are independent both within and across centers, the log-likelihood for the pooled dataset can be expressed as the sum of the local log-likelihood functions:
\begin{align}
\ell(\btheta) = \sum_{k=1}^K \ell_k(\btheta).    
\label{eq: loglik}
\end{align}
Let $\widehat\btheta$ be the (fictive) maximum likelihood estimator (MLE) of $\btheta$ based on $\bD$, which is obtained by maximizing the log-likelihood function $\ell(\btheta)$ with respect to $\btheta$, and let $\widehat\btheta_k$ be the MLE based on $\bD_k$ in center $k$.

\subsection{Single-communication, second order approximation}
In order to derive the federated single-communication estimator that was derived in \cite{Jonker2024}, the log-likelihood function in center $k$ is written as a second order Taylor expansion around the local MLE $\widehat\btheta_k$ (under the assumption that the log-likelihood function is sufficiently smooth):
\begin{align*}
\ell_k(\btheta) = \ell_k(\widehat\btheta_k) + \tfrac{1}{2}(\btheta-\widehat\btheta_k)^\top \bH_k(\widehat\btheta_k) (\btheta-\widehat\btheta_k) + O_p(\|\btheta-\widehat\btheta_k\|_2^3),     
\end{align*}
with $\bH_k(\widehat\btheta_k)$ the Hessian matrix of the log-likelihood function in center $k$, evaluated in $\widehat\btheta_k$, and where $O_p(\|\btheta-\widehat\btheta_k\|_2^3)$ is a remainder term of the order $\|\btheta-\widehat\btheta_k\|_2^3$. In this expansion, the first-order term involving the gradient is absent because, by definition, $\widehat{\btheta}_k$ maximizes the local log-likelihood function, implying that this gradient equals zero. Replacing each local log-likelihood function $\ell_k(\btheta), k=1,\ldots,K$ in (\ref{eq: loglik}) by its second-order Taylor approximation yields
\begin{align*}
\ell(\btheta) &= \sum_{k=1}^K \Big\{\ell_k(\widehat\btheta_k)  + \tfrac{1}{2}(\btheta-\widehat\btheta_k)^\top \bH_k(\widehat\btheta_k) (\btheta-\widehat\btheta_k) + O_p(\|\btheta-\widehat\btheta_k\|_2^3)\Big\}\\
&= \Omega_2(\btheta) + \sum_{k=1}^K O_p(\|\btheta-\widehat\btheta_k\|_2^3),
\end{align*}
with 
\begin{align*}
\Omega_2(\btheta) = \sum_{k=1}^K \Big\{\ell_k(\widehat\btheta_k) + \tfrac{1}{2}(\btheta-\widehat\btheta_k)^\top \bH_k(\widehat\btheta_k) (\btheta-\widehat\btheta_k)\Big\}.
\end{align*}
Under the assumption of homogeneity across the centers, the local MLEs are expected to lie in a small neighborhood of the pooled data MLE $\widehat{\btheta}$. Since our objective is to find $\widehat{\btheta}$, we are primarily interested in values of $\btheta$ that are close to $\widehat{\btheta}$. Consequently, the remainder terms $O_p(\|\btheta-\widehat{\btheta}_k\|_2^3)$ are negligible relative to the second-order terms in the region of interest and can therefore be ignored. 

The value that maximizes $\Omega_2$ with respect to $\btheta$ is defined as the FI2 estimator and is equal to:
\begin{align*}
\widehat\btheta_{\FI 2} := \underset{\btheta}{\operatorname{argmax}} ~ \Omega_2(\btheta) = \Big(\sum_{k=1}^K \bH_k(\widehat\btheta_k)\Big)^{-1} \sum_{k=1}^K \bH_k(\widehat\btheta_k) ~\widehat\btheta_k,      
\end{align*}
(where FI stands for federated inference and the 2 refers to the order of the Taylor approximation).
In order to compute $\widehat\btheta_{\FI 2}$ at the coordinating server, each center transmits its local MLE together with the corresponding Hessian matrix evaluated at that estimate. Specifically, center $k$ shares
\begin{align*}
\big\{\widehat{\btheta}_k , ~\bH_k(\widehat{\btheta}_k)\big\}
\end{align*}
with the coordinating server. The second derivative of $\Omega_2$ with respect to $\btheta$ is equal to 
\begin{align*}
\bH_{\FI 2} := \sum_{k=1}^K \bH_k(\widehat\btheta_k).
\end{align*}    
It approximates the Hessian matrix of the global log-likelihood function. Consequently, $-\bH_{\FI2}$ provides an estimate of the Fisher information matrix. 

In \cite{Jonker2025}, it is shown that for  increasing local sample sizes the estimator $\widehat\btheta_{\FI 2}$ is asymptotically unbiased for $\btheta$, asymptotically Gaussian, and asymptotically efficient, implying that no information is lost as a result of performing inference without pooling the individual-level datasets. Since $-\bH_{\FI2}$ estimates the Fisher information matrix, its inverse can be used to estimate the asymptotic covariance matrix of $\widehat{\btheta}_{\FI2}$. Consequently, a two-sided $(1-2\alpha)100\%$ confidence interval for the $j$th component of $\btheta$, denoted by $\theta_j$, can be constructed as
\begin{align*}
\Big[\widehat{\theta}_{\FI2, j}-\xi_\alpha \sqrt{-(\bH_{\FI2})_{j,j}^{-1}} ~~;~~ \widehat{\theta}_{\FI2, j}+\xi_\alpha \sqrt{-(\bH_{\FI2})_{j,j}^{-1}}~\Big],     
\end{align*}
where $\xi_\alpha$ is the upper $\alpha$ quantile of the standard Gaussian distribution.

\subsection{Single-communication, third order approximation}
Although the estimator $\widehat{\btheta}_{\FI2}$ has attractive asymptotic properties, its finite-sample performance may be less accurate, particularly when the sample size is small. A possible explanation is that the second-order approximation fails to capture higher-order characteristics of the local log-likelihood functions, such as skewness. To improve finite-sample accuracy, we propose a new estimator based on a third-order Taylor approximation of each local log-likelihood function around its corresponding local MLE. 
To derive this estimator, the log-likelihood function in center $k$ is written as:
\begin{align*}
\ell_k(\btheta) &= \ell_k(\widehat\btheta_k) + \tfrac{1}{2}(\btheta-\widehat\btheta_k)^\top \bH_k(\widehat\btheta_k) (\btheta-\widehat\btheta_k)\\
&+ \tfrac{1}{6}\sum_{h,i,j=1}^{d,d,d} T_{k,hij}(\widehat\btheta_k)(\theta_h-\widehat\theta_{k,h})(\theta_i-\widehat\theta_{k,i})(\theta_j-\widehat\theta_{k,j}) +  O_p(\|\btheta-\widehat\btheta_k\|_2^4),     
\end{align*}
with $T_{k,hij}(\widehat\btheta_k)$ the third derivative of the log-likelihood function in center $k$ with respect to $\theta_h, \theta_i$ and $\theta_j$ (the coordinates of $\btheta$), evaluated in the local MLE $\widehat\btheta_k$. 
Define
\begin{align*}
\Omega_3(\btheta) &= \sum_{k=1}^K \Big\{\ell_k(\widehat\btheta_k)  + \tfrac{1}{2}(\btheta-\widehat\btheta_k)^\top \bH_k(\widehat\btheta_k) (\btheta-\widehat\btheta_k)\\
&~~~~+ \tfrac{1}{6}\sum_{h,i,j=1}^{d,d,d} T_{k,hij}(\widehat\btheta_k)~(\theta_h-\widehat\theta_{k,h})(\theta_i-\widehat\theta_{k,i})(\theta_j-\widehat\theta_{k,j})\Big\},
\end{align*}
and write
\begin{align*}
\ell(\btheta) = \Omega_3(\btheta) +  \sum_{k=1}^K O_p(\|\btheta-\widehat\btheta_k\|^4_2).
\end{align*}

The FI3 estimator for $\btheta$ is defined as the value that maximizes $\Omega_3(\btheta)$:
\begin{align*}
\widehat\btheta_{\FI 3} := \underset{\btheta}{\operatorname{argmax}} ~ \Omega_3(\btheta).      
\end{align*} 
To obtain this estimate, the derivative of $\Omega_3$ is set equal to zero and the estimate is found by solving this equation:
\begin{align*}
0 =
\sum_{k=1}^K
\Big\{
\sum_{j=1}^d
H_{k,rj}(\widehat{\btheta}_k)
(\theta_j-\widehat\theta_{k,j})
+
\tfrac{1}{2}
\sum_{i,j=1}^{d,d}
T_{k,rij}(\widehat{\btheta}_k)
(\theta_i-\widehat\theta_{k,i})
(\theta_j-\widehat\theta_{k,j})
\Big\}
\end{align*}
for $r=1,\ldots,d$.
There exist multiple numerical algorithms to do this. In the simulation studies described below, this is done by minimizing the sum (over $r$) of the squares of the right hand side in the previous display.

The second derivative of $\Omega_3(\btheta)$ with respect to $\btheta$ is a function of the unknown parameter $\btheta$. By replacing $\btheta$ by $\widehat\btheta_{\FI3}$, we obtain
\begin{align*}
\bH_{\FI3} := \sum_{k=1}^K
\left[
\bH_k(\widehat{\btheta}_k)
+
\left(
\sum_{j=1}^d
T_{k,hij}(\widehat{\btheta}_k)
(\widehat\theta_{\FI3,j}-\widehat\theta_{k,j})
\right)_{h,i=1}^{d,d}
\right],
\end{align*}
with $\widehat\theta_{\FI3,j}$ the $j$th coordinate of $\widehat\btheta_{\FI3}$.
The matrix $-\bH_{\FI3}$ is an estimate of the Fisher Information matrix and can be used for inference, e.g., the construction of confidence intervals or hypothesis testing.

To compute $\widehat{\btheta}_{\FI3}$ at the coordinating server, each center transmits its local MLE, the Hessian matrix, and the tensor of third-order derivatives, both evaluated at the local MLE. Specifically, center $k$ shares
\begin{align*}
\big\{\widehat\btheta_k, ~\bH_k(\widehat\btheta_k), ~\bT_k(\widehat\btheta_k)\big\}.
\end{align*}
Note that $\bT_k$ has dimension $d\times d \times d$, which will be large if $d$ is large.

\medskip

A Bayesian version was proposed in \cite{Jonker2024,Jonker2025}. To compute the estimator, the global log-posterior density is expressed in terms of the local log-posterior densities together with the global and local prior densities. The local log-posterior densities are then replaced by their second-order Taylor approximations around the corresponding local maximum a posteriori (MAP) estimators. Maximizing the resulting approximation with respect to $\btheta$ yields the Bayesian Federated Inference estimator of order 2 (BFI2). Analogously, replacing the local log-posterior densities by third-order Taylor approximations around the local MAP estimators and maximizing the resulting approximation with respect to $\btheta$ yields the Bayesian version of the FI3 estimator.

\subsection{Two-communications, second order approximations}
Suppose that a second communication round is allowed. One natural approach is to perform a single Newton–Raphson update to maximize the global log-likelihood, using $\widehat{\btheta}_{\FI2}$ (or $\widehat{\btheta}_{\FI3}$) as the initial value. Let $\bS(\btheta)$ and $\bH(\btheta)$ denote the gradient and Hessian of the global log-likelihood function based on $\bD$, respectively. The resulting updated estimator is then defined as
\begin{align*}
\widehat\btheta_{\FI2 +} &:= \widehat\btheta_{\FI2} - \big(\bH(\widehat\btheta_{\FI2})\big)^{-1} \bS(\widehat\btheta_{\FI2})\\
&~=\widehat\btheta_{\FI2} - \Big(\sum_{k=1}^K \bH_k(\widehat\btheta_{\FI 2})\Big)^{-1} \sum_{k=1}^K \bS_k(\widehat\btheta_{\FI 2})
\end{align*}
with $\bS_k$ the gradient and $\bH_k$ the Hessian of the log-likelihood function in center $k$. The equality in the previous display follows from independence of the data across the centers.
Note that $\bS_k(\widehat\btheta_{\FI 2})$ is not necessarily equal to zero, as the estimator $\widehat\btheta_{\FI 2}$ may be different from the MLE in center $k$. In order to compute $\widehat\btheta_{\FI2+}$, the coordinating server needs to share the estimate $\widehat\btheta_{\FI 2}$ with the local centers first, next the local centers compute the gradient and the Hessian of their local log-likelihood evaluated in $\widehat\btheta_{\FI 2}$, and return these to the coordinating server. Specifically, in the second communication round, center $k$ transfers
\begin{align*}
\big\{\bS_k(\widehat\btheta_{\FI2}), ~\bH_k(\widehat\btheta_{\FI2}) \big\}    
\end{align*}
to the coordinating server.

\subsection{Two-communications, third order approximations}
After computing $\widehat\btheta_{\FI 3}$, a Newton-Raphson step can also be performed as described in the previous subsection, but now with the starting value $\widehat\btheta_{\FI 3}$. As a more accurate alternative, approximate the global log-likelihood function by a third order Taylor expansion around $\widehat\btheta_{\FI 3}$. Like in the previous section the global gradient, Hessian, and third order derivatives can be written as a sum of the local ones. That means that this third order approximation can be computed if every local center shares the gradient, Hessian, and tensor of third-order derivatives evaluated in $\widehat\btheta_{\FI 3}$ with the coordinating server.
Specifically, center $k$ shares
\begin{align*}
\big\{\bS_k(\widehat\btheta_{\FI3}), ~\bH_k(\widehat\btheta_{\FI3}), ~\bT_k(\widehat\btheta_{\FI3})  \big\}.    
\end{align*}
Maximizing the obtained third order Taylor approximation yields an estimator which we denote as $\widehat\btheta_{\FI 3+}$.

\section{Simulation Studies}
\label{sec: simulation}
\subsection{Aim}
\begin{figure}[h]
\centering   
\includegraphics[scale=0.28]{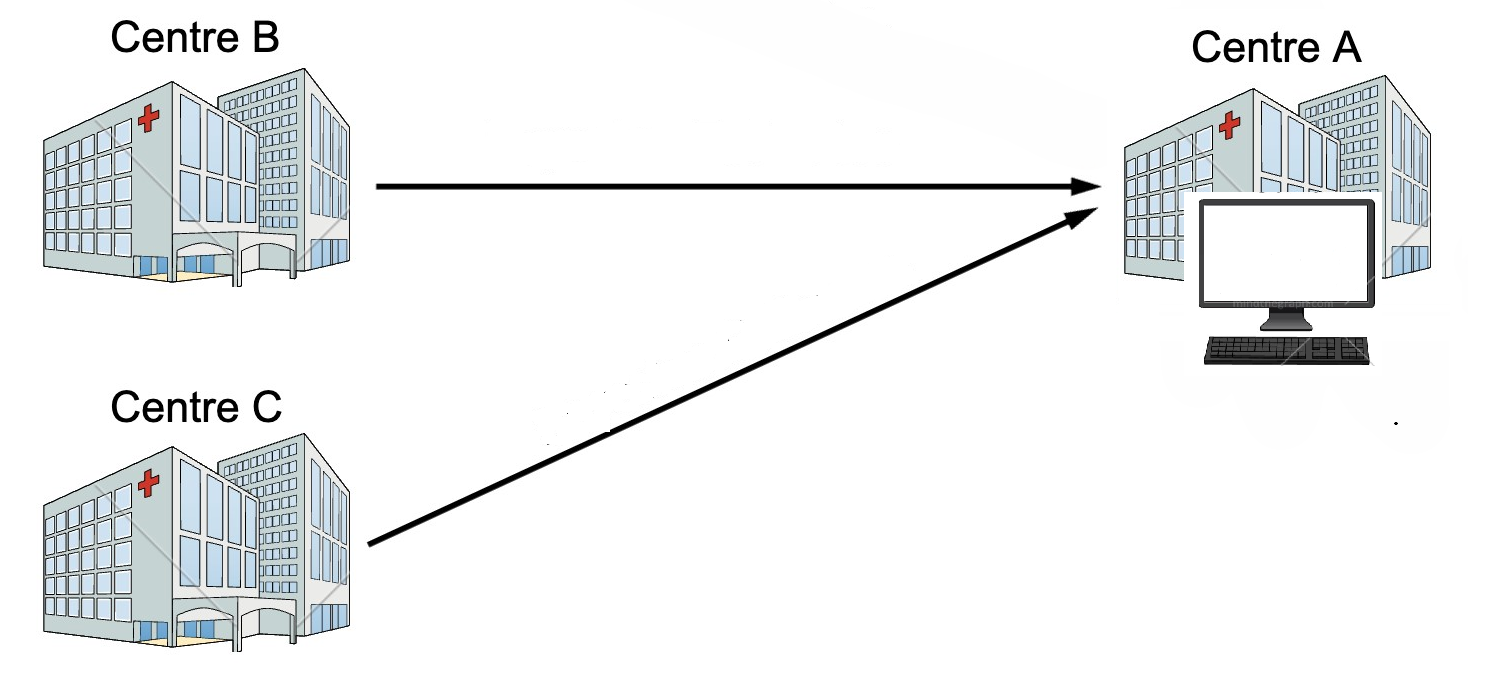}
\caption{Visualization of the setting in which ODAL2 operates. In center $A$ the MLE $\widehat\btheta_A$ is computed and shared with centers $B$ and $C$. Next these centers compute their local gradient and Hessian in $\widehat\btheta_A$ and transfer them to center $A$, where they are included in the analysis. The algorithms proposed in this paper can be applied in this setting, whereas the ODAL2 estimator can not be applied in the setting in Figure \ref{fig: BFI}.}
\label{fig: ODAL2}
\end{figure}

In this section the aim is to compare the performances of the estimators $\widehat\btheta_{\FI2},~ \widehat\btheta_{\FI3},~ \widehat\btheta_{\FI2+},~ \widehat\btheta_{\FI3+}$ and the estimator proposed in \cite{Duan}, which is called the ODAL2 estimator and will be denoted as $\widehat\btheta_{\ODAL2}$. The ODAL2 estimator is actually defined in a different setting as we work in. It is assumed that the coordinating server is located in one of the local centers, the lead center (see Figure \ref{fig: ODAL2}). That means that the data in the lead center are fully available for estimation and thus in total more information is available for estimation than in the setting we work in. How much extra depends on the sample size in the lead center and because of that it may become important which center is actually the lead center. For comparison we ignore the fact that we work in different settings, but the effect of the sample size in the lead center on the performance of $\widehat\btheta_{\ODAL2}$ will be considered. See \cite{Duan} for more information on the estimation strategy.

\subsection{Real data description}
\label{subsec:realdatadescription}
To be sure that simulations are run for realistic data settings, we use an existing dataset. 
The dataset that is used for the simulation study is a pooled dataset of 371 trauma patients from three hospitals. The binary variable mortality is used as an outcome ($Y=0$: survival, $Y=1$: death, mortality rate is 30\%), and the variables age, sex (0:males, 1:females), the Injury Severity Score (ISS, ranging from 1 (low) to 75 (high)) and the Glasgow Coma Scale (GCS, which expresses the level of consciousness, ranging from 3 (low) to 15 (high)) are used as the covariates in a logistic regression model. In order to better interpret the results, the continuous covariates were normalized in the pooled data. The three hospitals are a peripheral hospital without a neuro-surgical unit, a peripheral hospital with a neuro-surgical unit and an academic medical center. The sample sizes in the centers are respectively 49, 106 and 216. More details on the data are given in \cite{Jonker2024}. For every patient a vector of covariates $\bx$ and a binary outcome $y$ is observed. They are seen as a realisation of $(\bX, Y)$, which are related via
\begin{align}
\Pr(Y=1\mid \bX) = \frac{\exp(\btheta^T \bX)}{1+\exp(\btheta^T \bX)},    
\label{eq: logit}
\end{align}
where $\btheta$ is the vector of unknown regression parameters that needs to be estimated. 

The log-likelihood function for a dataset equals the sum of the log-likelihood functions of the individual patients. The contribution for a patient with data $(y,\bx)$ is given by
\begin{align*}
\log p(y,\bx) = \log p(y\mid \bx) + \log p(\bx),
\end{align*}
where $p(y\mid \bx)$ is the conditional probability of observing $y$ given $\bx$ and depends on the parameter vector $\btheta$ through the logistic regression model in (\ref{eq: logit}), whereas $p(\bx)$ denotes the density of the covariate vector, which is assumed not to depend on $\btheta$. Consequently, both the local and global log-likelihood functions can be decomposed into two components: one involving the conditional distribution of $y$ given $\bx$, which depends on $\btheta$, and one involving the distribution of the covariates, which does not. Since our objective is to estimate $\btheta$ rather than the covariate distribution, the latter component is constant with respect to $\btheta$ and can therefore be omitted from the analysis without loss of generality. Accordingly, we focus exclusively on the conditional log-likelihood $\log p(y\mid\bx)$.

\subsection{Simulation procedure}
\label{subsec:simulationprocedure}
The methodology presented in the paper, as well as the ODAL2 algorithm, is defined under a homogeneous setting, assuming that there are no systematic differences between the participating centers. To avoid model misspecification, the outcome for each patient is simulated from a logistic regression model conditional on that patient's covariates. Specifically, first the regression parameters in a logistic regression model for the pooled dataset are estimated, these estimates are seen as the true values $\btheta_0$ for the simulation procedure. Next for every patient, the mortality outcome is simulated given its covariates from a logistic regression model with the regression parameters $\btheta_0$. Thereafter, the patients are randomly assigned to one of the three  hospitals keeping the sample sizes in the hospitals fixed. In order to have an equal splitting in terms of male and female, the subgroups have been made with a stratification on sex. In a second simulation study the data are split up in more datasets to evaluate the performance in case of multiple small centers. The whole simulation procedure is repeated $M$ times.

\subsection{Metrics}
\label{subsec:metrics}
The aim of the federated analyses methods is to find the estimates that would have been found if the data had been pooled before doing the analysis. For this reason the gold standard is this pooled data estimate. 
Let $\widehat\btheta_{\FI2}^{(m)},~ \widehat\btheta_{\FI3}^{(m)},~ \widehat\btheta_{\FI2+}^{(m)},~ \widehat\btheta_{\FI3+}^{(m)}$, and $\widehat\btheta_{\ODAL2}^{(m)}$ be the different estimates of $\btheta$ computed in the $m^{th}$ run. The estimate based on the pooled data in run $m$ is denoted as $\widehat\btheta_{\pool}^{(m)}$. To quantify the performance of the estimators, the mean square error (MSE) is computed for every estimator. For the FI2 estimator this is calculated as:
\begin{align*}
	\MSES_{\theta_j,\FI2}&=\frac{1}{M}\sum_{m=1}^{M}\big(\widehat \theta_{\FI2,j}^{(m)}-\widehat \theta_{\pool,j}^{(m)}\big)^2, \quad &  \MSES_{\btheta,\FI2} = \frac{1}{d}\sum_{j=1}^{d} \MSE_{\FI2,j}
\end{align*}
and in a similar way for the other estimators. If the MSE is small, then the estimates computed with the corresponding federated estimation method are very close to the pooled estimates, which is the goal of all federated regression approaches. Note that in every simulation run, the pooled data estimates may be different from $\btheta_0$.

A second aim is to evaluate the accuracy of the outcome predictions obtained using the different estimators relative to those based on the model fitted to the pooled data. To assess predictive performance, the dataset is randomly divided into a training set (90\%), which is used for model estimation, and a test set (10\%), which is used for evaluation. For each patient in the test set, with covariate vector $\bx$, the predicted probability of mortality is computed as $\exp(\btheta^\top\bx)/(1+\exp(\btheta^\top\bx))$, where $\btheta$ is replaced by the estimator under consideration. 

Let $R_k^{(m)}$ be the test set of patients in center $k$ in simulation round $m$, and let $N_{test}$ be the sample size of the patients in the combined testing dataset. This number is the same in each cycle. The MSE of the predictions from the model in which the regression parameters are estimated by the FI2 algorithm is defined as:
\begin{align}
	\MSES_{p,\FI2}=\frac{1}{M}\sum_{m=1}^{M}\sum_{k=1}^{K}\frac{1}{N_{test}}\sum_{i\in R_k^{(m)}}\big(\widehat p_{\FI2, k i}^{(m)}-\widehat p_{\pool, ki}^{(m)}\big)^2,
    \label{eq:MSEp}
\end{align}
where $\widehat p_{\pool, ki}^{(m)}$ is the prediction for patient $i$, in center $k$ and in simulation round $m$, from the model in which the regression parameters are estimated by the MLE in the pooled training dataset. The MSEs for the other estimators are defined in a similar way.

\subsection{Results: Three medical centers}
\label{subsubsec:homogeneity}
For the different estimators the MSE was computed (Table \ref{tab: MSE3centers}). For ODAL2 the MSE depends on the choice of the lead center. Therefore three values of $\MSES_{\btheta,\ODAL2}$ are given. From the table it can be concluded that among the one-communication procedures FI2 and FI3 clearly outperform ODAL2, independently of the choice of the lead center. If an extra communication round (two in total) is performed to compute $\widehat\btheta_{\FI2+}$ or $\widehat\btheta_{\FI3+}$, the MSE decreases even further. The performance of ODAL2 increases with increasing sample size in the lead center, which is as expected as the amount of available information for estimation increases. When center 1 ($n_1=49$) is the lead center, the ODAL2 algorithm is highly unstable, producing extremely large MSE values in some simulation replicates. Therefore, we report the 5\% trimmed mean of the MSE. Even so, the trimmed MSE remains large, indicating that the algorithm performs poorly and unstably in this setting.  
\begin{table} [H]
	\centering
	\textbf{}\\[0.2cm]
{\small
    \begin{tabular}{cccc|ccc}
         \multicolumn{4}{c|}{$\MSES_{\btheta,method}$} & \multicolumn{3}{c}{$\MSES_{\btheta,\ODAL2}$}\\
     $\FI2$ & $\FI3$ & $\FI2+$ & $\FI3+$ & $n_1=49$ & $n_2=106$ & $n_3=216$\\
    \hline
     0.0125 & 0.0034 & 0.0002 & 0.0000 & 3.0624 & 0.3861 & 0.0675\\
     \end{tabular} \\[8pt]
}
\caption{MSE for the different estimation methods with $M=1000$. The values $n_1, n_2,$ and $n_3$ denote the sample sizes in the lead center for the ODAL2 estimator. For ODAL2, when center 1 ($n_1=49$) is the lead center, a 5\% trimmed mean was used because of the presence of outliers. 
}
\label{tab: MSE3centers}
\end{table}

\subsection{Results: Small subsets}
\label{subsubsec:smallsubsets}
In a second simulation study the performance of the estimators is evaluated if the local sample sizes are small. Instead of randomly allocating the patients to three local centers, the data will be divided into more subgroups, with a minimum sample size per center of 20, and the size of the dataset in the lead center is $n_1=50$. The results are given in \autoref{tab:smallsub.msem_trimmed}. 


\begin{table} [H]
	\centering
	\textbf{}\\[0.2cm]
{\small
     \begin{tabular}{c|ccccc}
		$K$ & $\MSES_{\btheta,\FI2}$ & $\MSES_{\btheta,\FI3}$ & $\MSES_{\btheta,\FI2+}$ & $\MSES_{\btheta,\FI3+}$ & $\MSES_{\btheta,\ODAL2}$\\
        \hline
        4   & 0.0260 & 0.0066 & 0.0008 & 0.0000 & 9.6503 \\ 
        6   & 0.0612 & 0.0252 & 0.0034 & 0.0003 & 9.1675 \\
		8   & 0.1139 & 0.0526 & 0.0102 & 0.0017 & 7.6159 \\
		10  & 0.1709 & 0.1110 & 0.0202 & 0.0035 & 15.7150 \\
		12  & 0.2285 & 0.1592 & 0.0328 & 0.0073 & 32.9060\\
	\end{tabular}\\[8pt]
}
	\caption{MSE with $M=1000$. The minimum local sample size is $20$ and size of the dataset in the lead center equals $50$. For ODAL2 the 5\% trimmed mean was compute. 
    }
	\label{tab:smallsub.msem_trimmed}
\end{table}
In every configuration, FI3 outperforms FI2, and FI3+ outperforms FI2+. Furthermore, all proposed methods outperform ODAL2, which exhibits considerable instability. As the number of centers, $K$, increases, the local sample sizes decrease, leading to higher MSEs.

Additional simulation studies were conducted using synthetic data with uncorrelated Gaussian covariates, with varying number of centers and sample size in the lead center. The results showed that when the data are extremely imbalanced such that the vast majority of observations are concentrated in the lead center, ODAL2 outperforms FI2. However, its performance still does not surpass that of FI3, FI2+ and FI3+.   

\begin{figure*}[t]
    \centering
    \includegraphics[width=\textwidth]{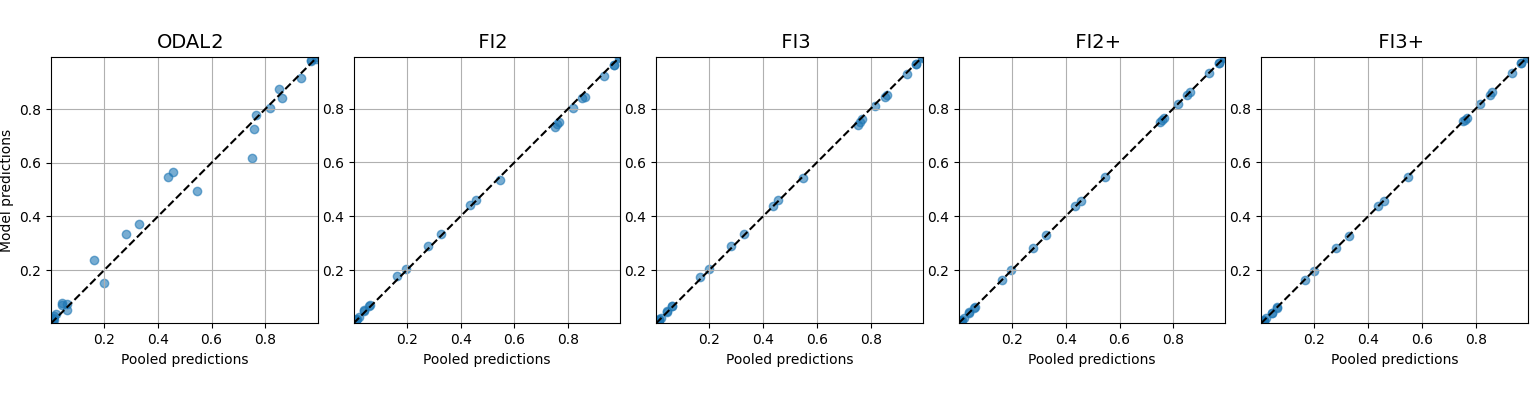}
    \caption{Scatter plots of the prediction probabilities in the test set, in a single run, and computed with the different methods (y-axis) versus the probabilities computed by the model that is estimated based on the pooled data. The dotted line is the $y=x$ line. The Pearson $R^2$ are 0.9890 (ODAL2), 0.9998 (FI2), 0.9999 (FI3), 1.0000 (FI2+), 1.0000 (FI3+).}
    \label{fig: probs}
\end{figure*}

Next, the MSE values are computed for the predictions obtained from formula \eqref{eq:MSEp}. The results are given in Table \ref{tab:smallsub.msep}. Figure \ref{fig: probs} shows scatter plots in which the predicted mortality probabilities in the test set computed with the different estimation methods are plotted against those based on the models estimated with the pooled data. The same conclusions apply for the predictions as for the parameter estimation.\\

\begin{table} [H]
	\centering
	\textbf{}\\[0.2cm]
    {\small
	\begin{tabular}{c|cccccc}
		$K$ &  $\MSES_{p,\FI2}$ & $\MSES_{p,\FI3}$ & $\MSES_{p,\FI2+}$ & $\MSES_{p,\FI3+}$ & $\MSES_{p,\ODAL2}$\\
        \hline
        4  & 0.0004 & 0.0001 & 0.0000 & 0.0000 & 0.0199\\
        6  & 0.0009 & 0.0003 & 0.0000 & 0.0000 & 0.0202 \\
		8  & 0.0018 & 0.0008 & 0.0001 & 0.0000 & 0.0204\\
		10 & 0.0031 & 0.0016 & 0.0002 & 0.0000 & 0.0201\\
		12 & 0.0045 & 0.0027 & 0.0004 & 0.0000 & 0.0207\\
	\end{tabular}\\[8pt]
    }
    \caption{MSE on predictions with $M=1000$. The minimum sample size of each local dataset is $20$ and for ODAL2 the sample size in the lead center is $n_1=50$.}
	\label{tab:smallsub.msep}
\end{table}

For relatively small local sample sizes, $\widehat{\btheta}_{\FI3}$, occasionally differs substantially from the pooled-data estimate. This is likely due to an unfavorable random allocation of observations across centers, combined with correlation between the covariates and the additional flexibility introduced by the third-order derivative term in the objective function. Although the resulting estimate $\widehat{\btheta}_{\FI3}$ may be aberrant, the corresponding predictions need not be, as they depend on the linear predictor $\widehat{\btheta}_{\FI3}^\top x$ rather than on the individual regression coefficients.
The estimator $\widehat{\btheta}_{\FI2}$ appears to be less sensitive. In practice, a simple diagnostic for detecting an aberrant FI3 estimate is to compare $\widehat{\btheta}_{\FI3}$ with $\widehat{\btheta}_{\FI2}$ for instance by assessing whether $\widehat{\btheta}_{\FI3}$ lies outside the 95\% confidence interval of $\widehat{\btheta}_{\FI2}$, in which case one may consider to replace it by $\widehat{\btheta}_{\FI2}$. This procedure was applied in the simulation study. It reduced the $\MSES_{\btheta,\FI3}$ to 0.104 ($K=6$), 0.0475 ($K=8$), 0.0866 ($K=10$), and 0.1353 ($K=12$) in Table \ref{tab:smallsub.msem_trimmed} and to 0.0027 in Table \ref{tab: MSE3centers}.  

\section{Discussion}
In this paper, we present several privacy-preserving algorithms for statistical inference in a federated setting that require either one or two rounds of communication between the participating centers and the coordinating server. The proposed methods are applicable to any parametric model, and their performance was evaluated through simulation studies for the logistic regression model. Although the algorithms were developed from a frequentist perspective, they can be naturally extended to the Bayesian framework, like in \cite{Jonker2024}.

The one-shot methodology based on second-order approximations of the local log-likelihood functions (or the log-posterior in a Bayesian setting) was previously presented in \cite{Jonker2024, Jonker2025}. To improve the estimation accuracy, we introduced an estimator based on third-order Taylor approximations.  By incorporating third-order information, the proposed approach captures local asymmetry of the log-likelihood or log-posterior that is neglected by quadratic approximations, resulting in more accurate estimates in the simulation studies.


Throughout this paper, we assumed a homogeneous setting in which all participating centers share the same underlying parametric model. Nevertheless, the proposed methodology can be extended to heterogeneous settings in a manner analogous to the second-order Bayesian Federated Inference framework described in \cite{Jonker2025}. In contrast, extending ODAL2 to heterogeneous settings is less straightforward because the initial parameter estimate is computed at a designated lead center before being communicated to the remaining sites. 

Future work will focus on translating the proposed methodology to a Bayesian setting and implementing it in the existing R package {\text BFI} that was developed for the Bayesian Federated Inference methodology for generalized linear models and parametric survival models as described in \cite{Jonker2024,Jonker2025,Massa,Pazira}. 

\section{Conclusions}
Including the third-order term in the Taylor approximation of the log-likelihood seems to improve the estimation accuracy of the aggregated estimator. This improvement has been demonstrated for logistic regression models in multiple settings. Although the methodology was developed from a frequentist perspective, it naturally extends to Bayesian inference and can be further generalized to heterogeneous federated learning settings.


\bigskip

\noindent
{\bf {Conflicts of interest}}\\
The authors declare no conflicts of interest.

\bigskip

\noindent
{\bf {Data availability}}\\
The Trauma dataset used in the simulation study is available in the R-package BFI.

\bigskip

\noindent
{\bf {Funding}}\\
The authors have nothing to report.

\bibliographystyle{unsrtnat}

\begin{thebibliography}{99}



\bibitem{Duan}
Duan R, Boland MR, Liu Z, Liu Y, Chang HH, Xu H, Chu H, Schmid CH, Forrest CB, Holmes JH, Schuemie MJ, Berlin JA,  Moore JH, Chen Y. Learning from electronic health records across multiple sites: A communication-efficient and  privacy-preserving distributed algorithm. Journal of the American Medical Informatics Association, 27(3), 376–385,  2019. https://doi.org/10.1093/jamia/ocz199

\bibitem{Jonker2024}
Jonker MA, Pazira H, Coolen ACC. Bayesian federated inference for estimating statistical models based on non‐shared multicenter datasets
Statistics in Medicine 43 (12), 2421-2438

\bibitem{Jonker2025}
Jonker MA, Pazira H, Coolen ACC. Bayesian Federated Inference for regression models based on non-shared medical center data Research Synthesis Methods 16 (2), 383-423

\bibitem{Jordan}
Jordan MI, Lee JD, Yang Y. Communication-Efficient distributed statistical inference. Journal of the American Statistical Association, 114(526), 668–681, 2018. https://doi.org/10.1080/01621459.2018.1429274

\bibitem{Kommusaar}
Kommusaar J, Elunurm S, Chomutare T, Kangasniemi M, Salanterä S, Peltonen L. A roadmap for federated learning  projects using health data to guide sustainable artificial intelligence development in the European Union. International Journal of Medical Informatics, 208, 106242, 2025. https://doi.org/10.1016/j.ijmedinf.2025.106242

\bibitem{Lassche}
Lassche G, Van Boxtel W, Ligtenberg MJ, Van Engen-Van Grunsven AC, Van Herpen CM. Advances and challenges in  precision medicine in salivary gland cancer. Cancer Treatment Reviews, 80, 101906, 2019. DOI: 10.1016/j.ctrv.2019.101906

\bibitem{Massa}
Massa E, Jonker MA. Bayesian Federated Inference with Systematically Missing Features. 3rd International Conference on Federated Learning Technologies and Applications (FLTA), 340-347, 2025. DOI: 10.1109/FLTA67013.2025.11336685

\bibitem{Luo}
Luo C, Islam Md. N, Sheils NE, Buresh J, Reps J, Schuemie MJ, Ryan PB, Edmondson M, Duan R, Tong J, Marks-Anglin A,  Bian J, Chen Z, Duarte-Salles T, Fernández-Bertolín S, Falconer T, Kim C, Park RW, Pfohl SR, Shah NH, Williams AE, Xu H,  Zhou Y, Lautenbach E, Doshi JA, Werner RM, Asch DA, Chen Y. Dlmm as a lossless one-shot algorithm for collaborative  multi-site distributed linear mixed models. Nature communications, 13(1): 1678, 2022.  https://doi.org/10.1038/s41467-022-29160-4

\bibitem{Mcmahan}
McMahan B, Moore E, Ramage D, Hampson S, Arcas BAY. Communication-efficient learning of deep networks from  decentralized data. Proceedings of the 20th International Conference on Artificial Intelligence and Statistics (AISTATS),  1273–1282, 2017. https://proceedings.mlr.press/v54/mcmahan17a/mcmahan17a.pdf

\bibitem{Mitani}
Mitani AA, Haneuse S. Small Data Challenges of Studying Rare Diseases. JAMA Network Open. 2020.  doi:10.1001/jamanetworkopen.2020.1965

\bibitem{Monaco}
Monaco L, Zanello G, Baynam G, Jonker AH, Julkowska D, Hartman AL, O’Connor D, Wang CM, Wong-Rieger D, Pearce  DA. Research on rare diseases: ten years of progress and challenges at IRDiRC. Nat Rev Drug Discov 21, 319-320, 2022.  doi: https://doi.org/10.1038/d41573-022-00019-z

\bibitem{Pazira}
Pazira H, Massa E, Weijers JAM, Coolen ACC, Jonker MA. Bayesian federated inference for survival models   Journal of Applied Statistics 53(2), 203-223, 2026. https://doi.org/10.1080/02664763.2025.2511932 

\bibitem{Rieke}
Rieke N, Hancox J, Li W, Milletari F, Roth HR, Albarqouni S, Bakas HR, Galtier MN, Landman BA, Maier-Hein K, Ourselin  S, Sheller M, Summers RM, Trask A, Xu D, Baust M, Cardoso MJ. The future of digital health with federated learning, NPJ  Digital Medicine, 3, 119, 2020. DOI: 10.1038/s41746-020-00323-1

\bibitem{Xiong}
Xiong R, Koenecke A, Powell M, Shen Z, Vogelstein JT, Athey S. Federated causal inference in heterogeneous 
 observational data, Statist. Med.; 42: 4418–4439, 2023. https://doi.org/10.1002/sim.986

\end{thebibliography}

\end{document}